\documentclass[onecolumn,preprintnumbers,amsmath,amssymb]{revtex4}
\usepackage{graphicx,subfigure,epsfig,epstopdf}
 \usepackage{graphicx}% Include figure files
\usepackage{dcolumn}% Align table columns on decimal point
\usepackage{bm}% bold math
\usepackage{amsmath,amssymb}
\input{epsf}
\input{epsfx}
\newcommand{\Dp}[2]{ \frac{\partial {#1}}{\partial {#2}} }

\newcommand{\D}[2]{\frac{\mbox{d} {#1}}{\mbox{d}{#2}}}

\newcommand{\fr}[2]{\frac{{#1}}{{#2}}}

\newcommand{\unit}[1]{\hspace{-1.3pt} {#1}}

\newcommand{\msc}[2]{{#1}{\mbox{\small{{#2}}}}}

\begin{document}
\title{Low-Threshold Surface-Passivated Photonic Crystal Nanocavity Laser}
%\author{Dirk Englund, Andrei Faraon, Bingyang Zhang, Yoshihisa Yamamoto \& Jelena Vu\v{c}kovi\'{c}}

%\affiliation{ Ginzton Laboratory, Stanford University, Stanford CA 94305}
%\date{Jan 29, 2007}

\author{Dirk Englund}
%\email{cvitae.org/englund/}
\altaffiliation{Authors contributed equally.}
\affiliation{Ginzton Laboratory, Stanford University, Stanford CA 94305}
\author{Hatice Altug}
\altaffiliation{Authors contributed equally.}
\affiliation{Electrical and Computer Engineering Department, Boston University, Boston MA 02215}
\author{Jelena Vu\v{c}kovi\'{c}}
\affiliation{Ginzton Laboratory, Stanford University, Stanford CA 94305}

%\section{Introduction}
\begin{abstract}
The efficiency and operating range of a photonic crystal laser is improved by passivating the InGaAs quantum well (QW) gain medium and GaAs membrane using an (NH$_4$)S treatment.  The passivated laser shows a four-fold reduction in nonradiative surface recombination rate, resulting in a four-fold reduction in lasing threshold.  A three-level carrier dynamics model explains the results and shows that lasing threshold is as much determined by surface recombination losses as by the cavity quality factor ($Q$). Surface passivation therefore appears crucial in operating such lasers under practical conditions.
\end{abstract}
%\pacs{42.50.Ct, 42.50.Dv, 42.70.Qs, 78.67.Hc}
%FIG1.pdf
\maketitle
\clearpage

%\section{Introduction}
Photonic crystals (PCs) allow unprecedented control over the radiative properties of integrated emitters. By defining small mode-volume, high-quality factor ($Q$) cavities in PCs, enhanced light-matter interaction becomes possible.  This property has opened possibilities in fields including cavity quantum electrodynamics, detection, and light sources. Lasers in particular stand to gain through dramatically improved lasing threshold, modulation rate, cost, and large-scale device integration. From their first demonstration\cite{Painter99science}, PC lasers have most commonly relied on QWs for optical gain.  However, QWs limit PC laser performance in many material systems because of large nonradiative (NR) surface recombination. This problem is particularly damaging in PC structures where embedded QWs expose a large surface area.  Here we address the NR recombination problem by surface passivation.  We show that (NH$_4$)S-mediated surface passivation of PC laser structures lowers the NR recombination rate by more than $4\times$ and leads to $4\times$ reduction of lasing threshold.  The increased efficiency extends the operating range from cryogenic to practical regimes, enabling room-temperature lasing at THz-modulation rates, as described in \cite{Englund2007APL_2}.  A three-level rate equations model fits our experimental data well and suggests that surface passivation is crucial for PC lasers in InGaAs/GaAs and other material systems with fast NR surface recombination. 
%, prompting many researchers to switch to quantum dot gain media\cite{Ellis2007APL, Nomura2006OpEx}

The PC nanocavity lasers consist of 172 nm-thick GaAs slabs patterned with 9x9 arrays of single-hole cavities defined in a square-lattice PC, similar to those described in Ref.\cite{Altug2006Nature}. A central stack of four 8-nm In$_{0.2}$Ga$_{0.8}$As QWs, spaced by 8-nm GaAs barriers, forms the gain medium.

This sample is passivated using a solution of 7\% (NH$_4$)S in water.  The treatment removes contamination and oxides from the GaAs and In$_{0.2}$Ga$_{0.8}$As surfaces and caps the fresh surface with sulfur atoms \cite{Oigawa1991JJAP}.  Samples were first cleaned in Leksol, acetone, and ethanol, then submerged in the (NH$_4$)S solution for 15 minutes at 35$^\circ$C, and finally air-dried, following Ref.\cite{Petrovykh2002SS}.  We measured the radiative and NR properties, as well as lasing characteristics, before and after surface passivation.  

Before presenting the experimental results, we describe the carrier dynamics at low temperature ($\sim10$K) using a three-level rate model.  Letting $N_E$ represent the pump level carrier concentration (populated above the GaAs-bandgap using a laser with power $L_{in}$), $N_G$ the QW lasing level carrier concentration (resonant with the cavity frequency), and $P$ the coupled cavity photon density, we have\footnote{$V_a$: pump active volume;  $\omega_p$: cavity angular frequency; $\tau_p=Q/\omega_p$: cavity ring-down time;  $G(N)$: gain; $\Gamma\approx 0.16$: gain confinement factor for cavity mode with $4\times$ 8nm QWs; $\eta$: pump energy absorption ratio; $\tau_r$: SE lifetime in unpatterned QW; $\tau_{PC,nr}$: NR lifetime in PC; $\tau_{E,f},\tau_{E,r},\tau_{E,nr}$: lifetimes of pump-level relaxation, SE, and NR transitions.}
\begin{eqnarray}
\label{eq:laser_rate}
\D{P}{t} &=& \Gamma G(N_G) P + \fr{F_{cav} N_G}{\tau_{r}} - \fr{P}{\tau_p} \\ \nonumber
\D{N_G}{t} &=& \fr{N_E}{\tau_{E,f}} - N_G \left( \fr{F_{cav} +F_{PC}}{\tau_r}+\fr{1}{\tau_{PC,nr}} \right) - \Gamma G(N_G) P \\ \nonumber
%\D{N_E}{t} &=& \eta \fr{\msc{L}{in}}{\hbar \omega_p V_a}  - N_E \left( \fr{1}{\tau_{E,r}}+\fr{1}{\tau_{E,nr}+\fr{1}{\tau_{E,f}} \right)
\D{N_E}{t} &=& \eta \fr{\msc{L}{in}}{\hbar \omega_p V_a}  - N_E \left( \fr{1}{\tau_{E,r}}+\fr{1}{\tau_{E,nr}}+\fr{1}{\tau_{E,f}} \right)
\end{eqnarray}

 In the center equation, the total lasing-level decay rate $d N_G/dt$ is separated into cavity decay, PC leaky-mode decay, and NR loss: $1/\tau_{G} = (F_{cav} +F_{PC})/\tau_r + 1/\tau_{PC,nr}$.  Here, $F_{PC}\approx 0.2$ expresses spontaneous emission (SE) rate quenching inside the PC bandgap compared to the SE rate $1/\tau_r$ in the bulk QW (following simulations in \cite{Englund05PRL}), while $F_{cav}$ denotes the SE rate enhancement into the lasing mode.%, so only the $F_{cav}/\tau_r$ in the top equation drives the cavity.  
 %Here, $V_a$ is the pump active volume of the laser cavity resonating at $\omega_p$, $\tau_p=Q/\omega_p$ is the cavity ring-down time, $\Gamma\approx 0.16$ the confinement factor, $G(N)$ the gain, $\eta$ the pump absorption ratio, and $\tau_r,\tau_{PC,nr},\tau_{E,f},\tau_{E,r},\tau_{E,nr}$ denote the lifetimes for SE in the unpatterned QW, for NR recombination in the PC, and for pump-level relaxation, SE, and NR transitions.
%PC bandgap: a/l=0.32-0.34  -- so if cavities were designed for a/l=0.33 at l=980 nm, then bandgap would only be ~ 90 nm wide -- not large enough

\begin{figure}%[htbp]
 \includegraphics[width=3in]{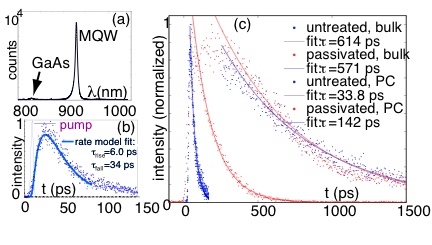} 
    \caption{{\footnotesize Low-temperature photoluminescence measurements on unpatterned and PC regions.  (a) PL from the bulk sample (after passivation). (b) Expanded view of PL from untreated PC region shows short lifetime $\tau_{PC} \approx 33.8$\unit{ps}; data is fitted by the rate model of Eqs.\ref{eq:laser_rate}.  (c) PL measurements for the untreated (blue) and passivated (red) samples, from the PC and unpatterned regions.}}
    \label{fig:Fig2}
\end{figure}
%XX FIT fit -- number cut off

We estimate the lifetime constants in Eqs. \ref{eq:laser_rate} from time-resolved photoluminescence (PL) recorded with a streak camera (Hamamatsu N5716-03) in a confocal microscope setup \cite{2007.OpEx.Englund}. The measurements are performed separately on PC mirrors and bulk regions with 3.5 ps-long excitation pulses at 780\unit{nm} wavelength and 82 MHz repetition rate (Fig.\ref{fig:Fig2}(b,c)). Samples were cooled to 10 K in a liquid-helium continuous-flow cryostat so that both unpassivated and passivated samples could be brought into lasing for comparison.  From a fit of Eqs.\ref{eq:laser_rate} to the rise-time of PL from the untreated sample, shown in Fig.\ref{fig:Fig2}(b), we estimate the relaxation time from the pump level into the lasing level at $\tau_{E,f}\sim$6\unit{ps}. The passivated sample also gives $\tau_{E,f}\sim$6\unit{ps}.  %Finally, we assume the pump-level to be dominated by relaxation into the QW at rate $1/\tau_{E,f}$ since radiative decay from the GaAs bandgap is 30$\times$ weaker than for the QW, as shown in Fig.\ref{fig:Fig2}(a). %(XX what about $\tau_{nr}$??)% and non-radiative decay appears insignificant since the PL lifetime at the GaAs bandgap is not shorter in the PC than in the unpatterned region (XX question: why is PL ??). %Additional decay measurements of PL at the GaAs bandgap (at 817\unit{nm} at 10K) indicate a longer decay time in the patterned than in the unpatterned regions of the sample (185\unit{ps} vs. 185\unit{ps}) (due to somewhat reduced local density of electromagnetic states in the PC at this wavelength XXX), indicating that non-radiative recombination in the GaAs material is not signficant, and we therefore assume that $\tau_{E,nr} \gg \tau_{E,f}$.

Fig.\ref{fig:Fig2}(c) shows the reduction in NR surface recombination after passivation: the PL decay lifetime from the PC mirror region is extended to $\tau_{PC}\sim142$ \unit{ps} from $\tau_{PC}\sim 33.8$\unit{ps} before passivation, while the decay lifetime from the bulk QW has nearly unchanged lifetime $\tau_{bulk}\sim 571-614$\unit{ps} at 10$\mu$W pump power. This data is analyzed using the bottom two equations of Eqs. \ref{eq:laser_rate} applied to PC and bulk regions, i.e., $1/\tau_{i} =1/\tau_{i,nr}+F_{i}/\tau_r$ with $i$ denoting bulk or PC ($F_{bulk}=1$).  Assuming $\tau_{nr,bulk}\gg \tau_r$, the lifetime data then lets us estimate the unpatterned bulk SE lifetime $\tau_{r}\approx$ 654 (605)\unit{ps} and NR lifetime $\tau_{PC,nr} \approx$ 35.5 (149)\unit{ps} in the PC mirrors before (after) passivation.  We assume equal NR lifetime across the cavity and surrounding PC mirrors since the diffusion length of rate-limiting holes $\sim 3\mu$m, greatly exceeding the cavity size. %The results are summarized in Table \ref{tab:lifetimes}.  
%On these time-scales, the lasing-level QW carrier density $N_G$ can be assumed to follow $N_E$ directly since the relaxation time $\tau_{E,f}\sim 6$\unit{ps} is much shorter than the other time scales.  Now we use Eqs.\ref{eq:laser_rate} in the bulk and in the patterned regions (without cavity, $p=0$), to obtain 

%\begin{eqnarray}
%\label{eq:rates}
%\fr{1}{\tau_{bulk}} &=&  \fr{1}{\tau_{r}}  + \fr{1}{\tau_{bulk,nr}} \\ \nonumber
%\fr{1}{\tau_{PC}} &=&  \fr{F_{PC}}{\tau_{r}}  + \fr{1}{\tau_{PC,nr}} 
%\end{eqnarray}

To put this reduction in NR loss rate into perspective and compare it to reports on other types of structures, we extract the surface recombination velocity $S$ that describes the recombination at the QW surface. From the lifetime data in Fig.\ref{fig:Fig2}(c), it is clear that most NR recombination results from the PC holes.  The effect of passivation is therefore to reduce $S$, and a simple model allows us to quantify by how much (pump power is small enough to neglect Auger recombination). The diffusion and recombination of the QW carrier concentration $N_G$, uncoupled to the PC cavities, are described by the equation (following \cite{Hayes1988APL})% (Auger and radiative recombination become more important loss mechanism at higher pump powers)%\footnote{$\tau_{nr,PC}^{-1}=\tau_{nr,PC,bulk}^{-1}+\tau_{nr,PC,surf}^{-1}\approx \tau_{nr,PC,surf}^{-1}$}
\begin{equation}
\label{eq:diffusion}
\Dp{N_G}{t}=D \nabla^{2} N_G - N_G\fr{F_{PC}}{\tau_r},
\end{equation}
where $D$ is the ambipolar diffusion coefficient.  Surface recombination enters through the boundary condition $D \Dp{N_G}{r} + S N_G=0$.  Assuming isotropic minority-carrier density over the PC period $a=315$\unit{nm}, the total recombination rate of the PC depends only on the exposed QW surface area.  This area is equal if the PC is replaced with an array of mesas whose radius equals the PC hole radius $r$. Eq.\ref{eq:diffusion} is then easily solved in cylindrical coordinates\cite{Hayes1988APL}, giving the total recombination rate $1/\tau_{PC}=F_{PC}/ \tau_{r}+1/\tau_{PC,nr}=F_{PC}/\tau_{r}+2 S/r$, i.e., $\tau_{PC,nr}=r/2S$.  We then find that $S\approx 1.7\cdot 10^{5}$\unit{cm/s} ($4.0 \cdot 10^{4} $\unit{cm/s}) for the original (passivated) structure. This value for the surface recombination velocity is somewhat lower than previous room-temperature measurements on similar InGaAs/GaAs structures by \cite{Wenzel2004SST,Hu1994JAP}, which put it at between $\sim 1\cdot 10^{5}$ and $5\cdot 10^{6}$\unit{cm/s}.  This is expected, since $S \propto v_{th}\approx \sqrt{3k T/m^{*}}$, the thermal velocity, which is $\sim 6\times$ smaller at 10K \cite{Sze1981}.  Our observation of a four-fold lowering in $S$ with surface passivation is similar to other reports with (NH$_4$)S \cite{Boroditsky2000JAP}.  However, better passivation results could probably be achieved with (NH$_4$)S$_x$, $x>1$, for which up to 50$\times$ improvement was reported \cite{Wenzel2004SST}.  %Our erlier assumption $a<D/S$ is justified with $D\approx 20 $\unit{cm$^{2}$/s} \cite{Hu1994JAP}. XXX put in later

%\begin{table}%[htdp]
%\label{tab:lifetimes}
%\begin{center}
%\begin{tabular}{l|c|c|c}
%  &$\tau_{r}$ (ps) & $\tau_{PC,nr}$ (ps) \\
% \hline
%original & 699.8 & 35.5 \\
%passivated & 577.5 &188.3 \\ %XXX put in better data  
%\end{tabular}
%\caption{Radiative SE lifetime in the unpatterned region, $\tau_r$ and NR lifetimes in the PC, $\tau_{PC,nr}$, before and after passivation.}
%\end{center}

%\end{table}

\begin{figure}%[htbp]
 \includegraphics[width=3in]{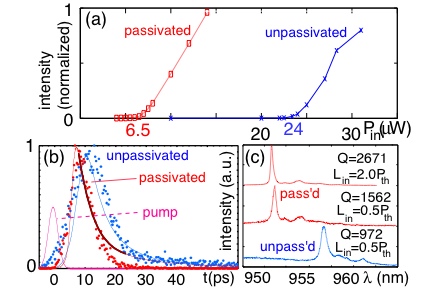}
     \caption{{\footnotesize Cavity resonances below and above lasing threshold.  (a) Lasing curves for unpassivated and passivated structures at low temperature (10K) with pulsed excitation (3.5\unit{ps}, 13\unit{ns}-rep.). Passivation reduces threshold from 24\unit{$\mu$W} to 6 \unit{$\mu$W} averaged power (measured before an objective lens focusing to a $\sim 3\mu$m-radius-spot). (b) Laser time response for untreated (blue) and treated (red) samples at 10K; Eqs.\ref{eq:laser_rate} fit the data well.  The treated laser shows an exponential decay time of 6.1\unit{ps} (thick fit). Some deviations at longer times are caused by background PL from regions not coupled to the resonant mode. (c) Cavity resonances below and above lasing.  Passivation lowers the resonance wavelength and slightly increases $Q$, as seen in the untreated (blue) and treated (red) cavities spectra at $1/2$ threshold pump power.  Top spectrum (red): lasing of passivated structure, pumped $2\times$ above threshold.  }}
      \label{fig:Fig4}
\end{figure}

With this understanding of the carrier dynamics in the PC, we now consider the coupled cavity array laser.  Microscope images show that only 7-9 cavities simultaneously lase in a single mode as fabrication inaccuracies lifted the cavity array's resonance degeneracies.  Fig.\ref{fig:Fig4}(c) shows that the passivation treatment slightly blue-shifts the cavity resonance and raises $Q$ by $\sim 1.5\times$ by cleaning and thinning the membrane, as observed in digital cavity etching \cite{Hennessy2005APL}.  The figure also shows the passivated structure when pumped $2\times$ above threshold; $Q$ is then raised to $2670$ due to gain.  We estimate the average SE enhancement factor $F_{cav}$ of emission coupled to the PC cavities from a lifetime measurement of the non-lasing cavity measured at $\sim 1/2$ lasing threshold, giving $\tau_{\mbox{\small{cav}}}\approx 17$\unit{ps}. The relation for the cavity-coupled SE rate, 
\begin{equation}
\fr{1}{\tau_{\mbox{\small{cav}}}} = \fr{F_{cav}+F_{PC}}{\tau_{r}}  + \fr{1}{\tau_{PC,nr}},
\end{equation}
gives $F_{cav}\approx 33$.  

Fig.\ref{fig:Fig4}(a) shows the lasing curves for the original and passivated structures and indicates a four-fold reduction in threshold. This reduction in the pump power $\msc{L}{in}$ directly follows from Eqs. \ref{eq:laser_rate}:  for threshold, we solve Eqs. \ref{eq:laser_rate} in steady-state with $P V_{mode}=1$ (an average of one photon inside the resonant mode) and $N_G\rightarrow N_{tr}$, the transparency carrier concentration where QW gain cancels absorption.  Neglecting the slow pump-level radiative recombination $\tau_{E,r}$, this gives \vspace{6pt} \\ 
$L_{in,th}=$
\begin{eqnarray}
\label{eq:laser_threshold}
 \fr{\hbar \omega_p}{\tau_p \eta} \fr{V_a}{V_{mode}} \left[ N_{tr} V_{mode} \left(F_{PC} \fr{ \tau_p}{\tau_r}+\fr{\tau_p}{\tau_{nr}} \right) + 1  \right]\left(1+\fr{\tau_{E,f}}{\tau_{E,nr}}\right)
\end{eqnarray}
For typical parameters, $N_{tr}\approx 10^{18}$\unit{cm$^{-3}$} \cite{1995Coldren} and $V_{mode} \approx 6 (\lambda/n)^{3}$, the first term in the brackets dominates.  Within this term, the nonradiative part $\tau_p/\tau_{nr}$ dominates the radiative one $F_{PC} \tau_p/\tau_r$.  Thus, in PC lasers using InGaAs QWs, or other gain media with similar surface recombination velocity, threshold is largely determined by NR recombination losses at the QW and GaAs membrane surfaces.  After passivation, Eq.\ref{eq:laser_threshold} predicts a threshold reduction by factor 4.1 of the original value if the NR pump-level loss rate $1/\tau_{E,nr}$ is assumed much smaller than the relaxation rate $1/\tau_{E,f}$ into the lasing level (otherwise an even larger reduction).  We measured a decrease by factor $3.7$, which is shows good agreement with the predicted value. The differential quantum efficiency, on the other hand, is nearly unaffected by the NR recombination rate, as can be easily derived from the rate equations (the physical reason is that once lasing begins, the stimulated emission rate is much faster than the NR loss rate.)   %, and we therefore believe that the passivation of the GaAs is not significant in this case

One of the most remarkable aspects of the PC nanocavity laser is the extremely fast modulation rate.  In Fig.\ref{fig:Fig4}(b), we present streak camera measurements of the lasing response to 3.4-ps-long pump pulses.  The low-temperature measurements for the passivated and unpassivated samples were obtained at the same average pump power of $\sim$28\unit{$\mu$W} (3.5\unit{ps}, 13 ns repetition), and the normalized lasing response is compared in the red and blue plots.  After passivation, the laser responds somewhat faster with an exponential decay time of $6.1$\unit{ps}.  We attribute this speed-up largely to relatively higher pump power above threshold (due to lower NR loss and higher cavity $Q$). Faster time response is possible at higher pump power, as noted in \cite{Altug2006Nature}. The rate model of Eqs.\ref{eq:laser_rate} explains these time-response measurements well, as shown in the continuous-line fits. 

%We measured significantly faster lasing response at room temperature, with the lasing response nearly following the pump pulse.  This speed-up is due to faster carrier relaxation at room temperature, as indicated in the photoluminescence response of the unpatterned QW at room temperature Fig.\ref{fig:Fig5}(b):  the rise-time is streak-camera limited to less than 1.8\unit{ps} -- significantly shorter than the low-temperature rise-time of $\sim 6$\unit{ps}.  This behavior is captured well by our model, which is combined here with a filter that takes into account the 3.2-ps response time (FWHM) of the streak camera.  The measured PC nanocavity lasing time is limited by the streak camera response; our model predicts that the underlying response time is in fact much faster, approaching $\tau_p = 0.8$\unit{ps} at 2$\times$ threshold pump power when pumped with 160-fs laser pulses and implying modulation rates in the THz regime.

In conclusion, we have demonstrated the threshold-lowering effect of surface-passivation treatment of InGaAs QWs in a PC coupled nanocavity array laser.  The 4-fold reduction of NR surface recombination lowers the threshold pump power to 27\% of its original value.  Our three-level laser model agrees well with experimental observations and shows that NR recombination strongly affects lasing when the NR loss rate is faster than the modified SE rate in the PC.  In this regime, reducing the NR surface recombination rate lowers the lasing threshold as much as lowering the cavity loss rate $1/\tau_p$ would, but has the advantage of not slowing lasing modulation rate.  Using a carrier diffusion model, we calculate a drop in the QW surface recombination velocity from $S\approx 1.7\cdot 10^{5}$\unit{cm/s} to $3.2 \cdot 10^{4} $\unit{cm/s} after passivation; comparing this to literature, we believe that our results could be improved by applying better surface passivation techniques \cite{Wenzel2004SST,Petrovykh2005}. The increased efficiency achieved in our lasers alleviates heating problems, which opens the door to room-temperature and CW operation \cite{Englund2007APL_2} and brings PC lasers closer to practical applications.  

The authors thank Dr. D.Y. Petrovykh for his helpful comments. This work was supported by the MARCO IFC Center, NSF Grants ECS-0424080 and ECS-0421483, the MURI Center  (ARO/DTO Program No.DAAD19-03-1-0199), as well as the NDSEG Fellowship (D.E.).   

\end{document}